\documentclass[runningheads]{llncs}
\usepackage{environ}
\usepackage{calc,amsmath,amssymb,amsfonts}
\usepackage{algorithmic}
\usepackage{mathtools}
\usepackage{graphicx}
\usepackage{textcomp}
\usepackage{xcolor}
\usepackage{tikz}
\usepackage{hyperref} 
\usepackage{tabularx}
\usepackage{boogie}
\usepackage{eiffel}
\usepackage{listings}
\usepackage{soul}
\usepackage{colortbl}
\usepackage{wrapfig}

\newcommand{\e}[1]{\mbox{\lstinline[basicstyle=\normalsize]|#1|}}

\usepackage[LGR,T1]{fontenc} 
\usepackage[greek,english]{babel}

\usepackage{cite}

\def\BibTeX{{\rm B\kern-.05em{\sc i\kern-.025em b}\kern-.08em
		T\kern-.1667em\lower.7ex\hbox{E}\kern-.125emX}}

\newcommand{\definitionlabel}[2]{%
	\expandafter\xdef\csname label@name@#1\endcsname{#2}%
	\label{def:#1}%
}

\newcommand{\referencename}[1]{%
	\csname label@name@#1\endcsname%
}
\newcommand{\comment}[2][]{
	\par
	\footnotesize
	-\hspace{0.1mm}-\textit{#2}
	\ifx\\#1\\%
	\else
	\definitionlabel{#1}{#2}%
	\fi
	\par
	\normalsize 
}

\newcommand{\rightcomment}[2][]{
	\par
	\raggedleft
	\comment[#1]{#2}
	\raggedright
}

\newenvironment{mytable}[1][\columnwidth]
{
	\noindent
	\renewcommand{\arraystretch}{1.1} 
	\begin{tabular*}{\columnwidth}{@{}l@{\hspace{0.5em}}l@{\hspace{0.5em}}l@{\extracolsep{\fill}}r@{}}
	}{
	\end{tabular*}
}

\newenvironment{definitionn}[3][0.66]{%
	\vspace{0.15em}\noindent
	\begin{minipage}[c]{(#1\columnwidth)}%
		\def\definitionend{%
		\end{minipage}%
		\hfill
		\begin{minipage}[c]{\dimexpr\columnwidth-(#1\columnwidth)-\columnsep\relax}%
			\raggedleft\footnotesize -\hspace{0.1mm}-\textit{#3}%
		\end{minipage}\par\vspace{0.5em}%
		\definitionlabel{#2}{#3}%
	}%
}{%
	\definitionend
}

\newenvironment{mydefinition}[2][]{
	\begin{definitionn}[0.66]{#1}{#2}
	}{
	\end{definitionn}
}

\begin{document}
\title{Loop unrolling: formal definition and application to testing}
%
%
\author{Li Huang\inst{1}\orcidID{0000-0003-3531-4045} \and
Bertrand Meyer\inst{1}\orcidID{0000-0002-5985-7434} \and
Reto Weber\inst{1}\orcidID{0009-0001-9262-4843}}
%
%
\institute{{Constructor Institute of Technology, Schaffhausen, Switzerland}
\url{https://institute.constructor.org}\\
\email{\textit{first}.\textit{last}@constructor.org}}
\maketitle              
\begin{abstract}
Testing processes usually aim at high coverage, but loops severely limit coverage ambitions since the number of iterations is generally not predictable. Most testing teams address this issue by adopting the  extreme solution of limiting themselves to \textit{branch coverage}, which only considers loop executions that iterate the body either once or not at all. This approach misses any bug that only arises after two or more iterations.

To achieve more meaningful coverage, testing strategies may \textit{unroll} loops, in the sense of using executions that iterate loops up to \textit{n} times for some \textit{n} greater than one, chosen pragmatically in consideration of the available computational power.

While loop unrolling is a standard part of compiler optimization techniques, its use in testing is far less common. Part of the reason is that the concept, while seemingly intuitive, lacks a generally accepted and precise specification. The present article provides a formal definition and a set of formal properties of unrolling. All the properties have mechanically been proved correct (through the Isabelle proof assistant).

Using this definition as the conceptual basis, we have applied an unrolling strategy to an existing automated testing framework and report the results: how many more bugs get detected once we unroll loops more than once? 

These results provide a first assessment of whether unrolling should become a standard part of test generation and test coverage measurement.

\end{abstract}
\section{Loops, coverage and unrolling} \label{intro}

 The issue addressed in this work, both theoretically and empirically, is a basic question of software testing: is branch coverage, the prime practical measure of testing effectiveness, justified in its drastic simplification of treating a loop like a conditional, whose body is executed either once or not at all, even though in an actual program run the body can be executed any number of times? Should we instead ``unroll'' loops, improving the approximation by considering not just zero or one but any number of iterations, up to a set limit? We will first define this concept of unrolling rigorously through a mathematical model. Then, using a set of practical examples and an automated test-generation framework relying on formal verification, we will examine how much (if anything) testing strategies miss when they limit themselves to standard branch coverage and, conversely, how many more bugs we can find if we unroll loops.

 \subsection{Loops} \label{loops}

A key property of computers is their ability to repeat operations, often many times. The corresponding construct in programming languages is the loop. (Functional languages use recursion or an equivalent mechanism instead, but this discussion assumes an imperative language.) In its general form, a typical loop L may be written \e{until} \e{e} loop B end, with the following execution behavior: evaluate \e{e} (a boolean expression); if its value is True, do nothing; if its value is False, execute \e{B} (the ``loop body'', an instruction or sequence of instructions) and repeat the entire process from the beginning. 
The loop can also be written \e{while c loop B end} where \e{c} is the logical negation of \e{e}. The two forms are equivalent; this discussion will stick to the \e{until} variant. It can be convenient to include an initialization clause with the keyword \e{from}.

A characteristic of such loop constructs is that it is impossible to predict statically (in other words, from the program text) how many iterations of \e{B} a particular execution of the loop will produce; different executions, with different input data, may result in different numbers of iterations. (If the program is buggy, the loop may also fail to terminate after a finite number of iterations.) This unpredictability is one of the key challenges of software verification, particularly automatic test generation and associated measures of test coverage.

 \subsection{Branch coverage} \label{branchcoverage}

Defined broadly, test coverage is a criterion assessing what share of a program's potential executions a given test suite (a set of tests for the program) exercises. The reason for defining coverage measures is that if we want to use testing to estimate the quality of the code, and more specifically the number of \textbf{remaining} bugs (as opposed to using testing just for finding bugs \cite{meyerprinciples}), we face the obvious obstacle that \textit{any} realistic program has an infinite or intractably large number of possible executions, forcing us to select
\cite{meyer_class_invariants}
a small subset of them --- the test suite --- for the test campaign; but we need to have some idea of how representative the test suite is of the full set. While many measures of test coverage have been proposed (see e.g. \cite{ammann2003coverage} for a survey), by far the most commonly used in industry is branch coverage, which measures the percentage of the program's possible \textit{control paths} (paths in the control flow of the program) being exercised. ``Achieving branch coverage'' means reaching 100\% of those possible paths; in practice, many development teams in industry set a lower percentage, such as 80\%, as the condition for shipping a product. While empirical studies have uncovered the limits of branch coverage, showing in particular \cite{branchcoverage} (see also \cite{gopinath} and \cite{jalali}) that an extensive testing campaign can reach a plateau at over 90\% coverage then continue to find bugs long after that stage, they have not affected the status of branch coverage as a key criterion in practical software development.

\begin{wrapfigure}{r}{3cm}
\centerline{{\includegraphics[width=2.3cm]{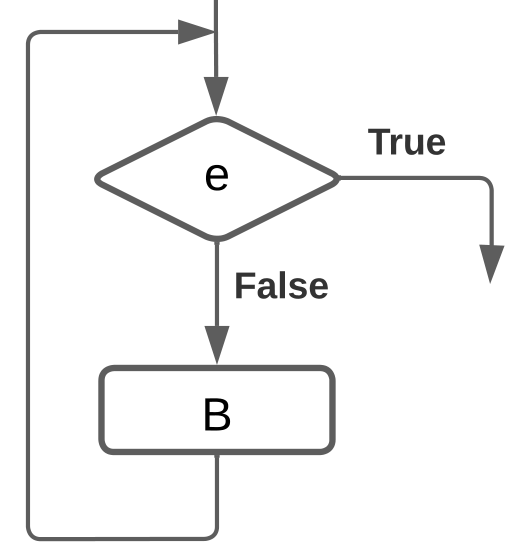}}
}
\caption{Control flow for a loop}
\label{figure: decision}
\vspace{-0.3cm}
\end{wrapfigure}

The definition of branch coverage used in practice makes a critical simplification with respect to loops. The obvious purpose is to skirt the major issue mentioned above, the impossibility of predicting how many times a loop will be iterated. The simplification is, however, drastic: branch coverage considers only two paths for a loop (Fig. \ref{figure: decision}), one which executes \e{B} once, and one that exits immediately. In other words, it reduces the loop \textbf{until} \e{e} \textbf{loop} \e{B} \textbf{end} to a simple conditional instruction \textbf{if not} \e{e} \textbf{then} \e{B} \textbf{end}.

In the reality of program execution, the set of possible paths is infinite, following the upward arrow of Fig. \ref{figure: decision} an arbitrary number of times \e{n $\ \geq \ $ 0}. As a proxy for actual executions, branch coverage misses cases in which \e{n} is 2 or more. Since loops are essential to computing, it is remarkable that such a brutal simplification has not prevented branch coverage from achieving in software development a role that industry massively finds essential. It is legitimate, however, to ask how much we may be losing by accepting this cavalier approach to loops; and, pragmatically, whether unrolling is feasible, and will enable us to find more bugs, the basic goal of testing.

The rest of this article develops answers to these questions. Section \ref{theory} introduces a mathematically precise definition of loop unrolling. Section \ref{implementation} explains how we added an automatic loop unrolling mechanism to an existing framework for generating tests automatically, relying on a combination of test and proof techniques. Section \ref{evaluation} analyzes and evalutes the results. Section \ref{related} reviews existing work, in particular in an area that is distinct from testing but closely related to it: model checking, which has introduced the unrolling-like notion of \textit{bounded} checking. Section \ref{threats} lists threats to validity and open issues.  Finally, section  \ref{conclusion} presents conclusions that current results suggest as to the suitability of adding loop unrolling to testing strategies and branch coverage measures. 

 \section{A mathematical definition of loop unrolling} \label{theory}

 \subsection{The need for a theoretical analysis} \label{why}

 The notion of loop unrolling is intuitively clear: when we need (for example for testing purposes) a finite approximation for a loop \textbf{until} \e{e} \textbf{loop} \e{B} \textbf{end}, with its potentially infinite set of possible executions, use a set of programs that execute \e{B} not at all, once, twice, three times and so on.

 In many cases this intuitive view is correct. For example with a loop computing the maximum of a non-empty array \e{a} indexed from 1 to \e{N}, \textbf{from} \e{M :=  -$\infty$; i := 1} \textbf{until} \e{i > N} \textbf{loop} \e{M := max (M, a [i]); i := i + 1} \textbf{end}, executing the body \e{k} times for \e{k < N} will yield the maximum of the array slice at indexes 1 to \e{k}, which is indeed an approximation of the final result. But sometimes this notion of approximation is far less clear. Assume for example that \e{M} is a positive integer, possibly large, and consider the loop   \textbf{from} \e{x :=  -M; i := 1} \textbf{until} \e{x > 0} \textbf{loop} \e{x := -M + i; i := -2 $*$ i} \textbf{end}. It yields \e{x = -M +  $2^j$} for the smallest odd integer \e{j} such that \e{$log_2$ (j) > M}. For lesser values of \e{j}, however, the value of \e{x} fluctuates widely, further off from the result (for even values) than the original approximation \e{M}! Unlike with the previous example, iterating the loop body an even number of times, less than the final number, does not give us an ``approximation'' of the result in any intuitive (as opposed to theoretical) sense.

 More generally, we should not let ourselves be fooled by the view (informally OK, but not literally true) that ``executing the loop \textbf{until} \e{e} \textbf{loop} \e{B} \textbf{end} means executing \e{B} 0 times, or 1 time, or any number of times''. Depending on the details of the loop, certain numbers of iterations --- such as, say, 4 iterations --- may not be possible at all. A better formulation talks about executing the body \textit{some} number of times (not \textit{all} possible numbers below the maximum if any). The rest of this section develops the mathematical theory providing the precise framework removing any ambiguity or potential confusion.
 
  The theory's underpinnings come from  classic work in denotational semantics \cite{stoy1977denotational} (see also \cite{meyer1990introduction} and, and \cite{meyer2015theory}) and abstract interpretation \cite{cousot2021principles}. The presentation is based on earlier work \cite{meyerloopunrolling} which, however, did not involve full proofs of properties, let alone machine-supported ones. All the formal properties stated in the present article have been proved and machine-checked using the Isabelle theorem prover \cite{isabelle1} and are publicly available. To facilitate cross referencing, every theorem stated below comes with a name, such as {/Concat\_station/} below, appearing in smaller font; the same name appears in the Isabelle files for the corresponding property and its proof. (Names in all upper case, such as \e{CONCAT\_DEF}, also appear in the Isabelle files; they denote definitions and hence do not require proofs.)  

 \subsection{Assumptions}
	
	We consider a loop L in the simple form given above:  \e{until e loop B end}. The informal semantics is the usual one: execute the instruction \e{B} (“body”) 0 or more times, stopping as soon as \e{e} (the “exit condition”) holds.	This discussion studies how we can --- in particular for testing purposes --- approximate L by a sequence of nested conditionals:

	\begin{definitionn}[0.5]{inap}{Inapplicable program}
		L$_0$ = \textbf{check False end}
	\end{definitionn}
	\begin{mydefinition}[iplus]{REC\_DEF}
		L$_{i+1}$ = \textbf{if} ${\lnot}$ e \textbf{then} B; L$_i$ \textbf{end}
	\end{mydefinition}
	
(Here an instruction \textbf{check} p \textbf{end}, where p is a Boolean property, has no effect if p has value true upon execution, and otherwise makes the entire program in which it appears inapplicable. It corresponds to \textbf{fail} in the trace set model of section \ref{traceset}. One can think of it in practice as causing a run-time crash but more abstractly it is simply an incorrect instruction. In the special case used here, \textbf{check False end}, defining L$_0$, is simply a program that is never applicable, regardless of the input. Another name for \textbf{check} in some verification formalisms is \textbf{assume} \cite{leino2010dafny}; \textbf{check False end} is essentially the same as Dijkstra's ``abort'' \cite{dijkstra1976discipline}.)

Subsequent L$_{i+1}$, for  $i \ {\geq}\ 0$, are defined for a growing set of possible inputs: those that, in each case, require at most $i$ executions of \e{B} before rendering \e{e} \textbf{True}. 

Why is it important to produce such a sequence of approximations? A number of applications exist, for example in compiler optimization, high-performance computing and software verification; the concrete impetus for the present study is to improve on  branch coverage. True ``path coverage'' would imply covering any number of executions of \e{B}, which is impossible in the general case; branch coverage goes to the other extreme of restricting that number to zero and one. Unrolling provides an intermediate solution: “execute” the loop a variable number of times (not just one), tuning the unrolling level in accordance with the testing needs and the available computing resources (since a higher level requires more testing time).

	\subsection{Notation: traces and states} \label{trace}
For the purpose of this discussion, the semantics of a program P is given by the set Traces (P) of its (finite) traces for any given input. In fact we identify P with its traces.
	
A trace x is a finite, non-empty sequence of program states written {\textless}$x_1$, $x_2$, …{\textgreater}.

In this definition, a program state is defined by the values of the program variables (in a general sense, which for an object-oriented program will include the whole heap) as well as a program location (the indication of which instruction of the program an execution is currently at). Intuitively it corresponds to what you see if you stop the program during execution and look at what the debugger tells you.

If $s$ is a state (an element in a trace) and $e$ is an expression, $s [e]$ is the value of $e$ in state $s$. For example if $s$ is the state resulting from executing the instruction sequence $x := 2$; $y := 5$ the value of $s [x * y]$ is 10.

The i-th element (state), of a trace $x$ is written $x_i$. Its length (number of elements) is written {\textbar}$x${\textbar}. Since traces are non-empty, $|x|>0$ and there always exists a first state, $x_1$, and a last state, written $x_L$. A trace is “stationary” if it has only one element, i.e. is of the form {\textless}$x_1${\textgreater}. (In this case $x_1$ is the same as $x_L$.) A stationary trace corresponds to an empty execution, which leaves the state unchanged.

A concatenation operator using the symbol “+” is available on traces; for example, {\textless}m, n{\textgreater} + {\textless}n, o, p{\textgreater} is {\textless}m, n, o, p{\textgreater}. As this example suggests, x + y is defined if and only if $x_L$ = $y_1$: the last element of the first operand must be the same as the first element of the second one. The common element (n in the example) appears only once  in the concatenation. The intuition behind this rule is that concatenating two program traces only makes sense if the first program ends in a state from which the second one can take over.

Formally, $x + y$ is defined as the trace $z$ of length {\textbar}$x${\textbar} + {\textbar} $y${\textbar} – 1 such that $z_i$ = $x_i$ for 1 ${\leq}$ $i$ \ ${\leq}$ {\textbar}$x${\textbar} and $z_i$ = $y_i$–{\textbar}$x${\textbar}+1 for {\textbar}$x${\textbar} + 1 ${\leq}$ $i$ ${\leq}$ {\textbar}$z${\textbar}. 

The “+” operator is associative and may be used for more than two operands, as long as they satisfy the requirement that the final state of every operand is the same as the initial state of the next.

\begin{mydefinition}[concatassoc]{/Concat\_assoc/} If x + y = z: \end{mydefinition}

\begin{definitionn}[0.71]{concatstation}{/Concat\_station/} \begin{itemize} 
		\item If x is stationary, then y = z; if y is stationary then x = z.  
\end{itemize} \end{definitionn}

\begin{definitionn}[0.73]{concatorder}{/Concat\_order/} \begin{itemize} 
\item We say that x is a \textbf{prefix} of z, written x {\textless}= z (the relation is a partial order). If y is not stationary then x is a \textbf{proper} prefix and we write x {\textless} z.
\end{itemize} \end{definitionn}
\vspace{-0.3cm}
\begin{itemize}
	\item We also say that z is an \textbf{extension} (resp. proper extension) of x. (y is a “suffix” of z but we do not need that notion.)
\end{itemize}
A \textbf{test} is a condition --- in other words, a Boolean expression --- on states.

A trace $x$ \textbf{satisfies} a test $v$ if $s [v]$ holds for some state $s$ in $x$. (Remember that a state includes a program location.)

\begin{mydefinition}[extensionstable]{/Extension\_stable/}
Theorem: if $x$ satisfies $v $ and $x\leq$$z$, then $z$ satisfies $v$.
\end{mydefinition}
	
\vspace{-0.3cm}
\subsection{Trace sets} \label{traceset}
\vspace{-0.1cm}

Instructions and programs (in the underlying programming language) will be defined by their \textbf{trace sets}. A trace set is what the name indicates: a set of traces.

\textbf{skip} is the set of stationary traces (those of the form {\textless}$x_1${\textgreater}, with just one state).

\textbf{fail} is the empty trace set. (Note that traces themselves cannot be empty, as they always have an initial state and a final state –-- which are the same for a stationary trace --– but a trace set can be empty.)

We say that a trace set A \textbf{tests} c, or is \textbf{a test of} c, if it contains a trace  satisfying c. 

If A and B are trace sets, A + B, also written A ; B, is the set of traces

\begin{mydefinition}[concatdef]{[CONCAT\_DEF]}
	$\{z \text{\ \textbar\ } \exists x:\text{A}, y:\text{B \textbar\ Z}  =x + y\}$
\end{mydefinition}

In other words, the set of all traces obtained by concatenating a trace from A and a trace from B. Unlike the “+” operator on traces, the “+” operator on trace sets is defined for any operands. (Non-concatenable trace pairs in A and B, meaning pairs such that $x_L \neq  y_1$, simply do not yield any element of A + B.) 

Theorems:

\begin{mytable}
	 \textbf{fail}&=& \textbf{fail} ; A & \comment[concatfail1]{/Concat\_fail1/}\\
	              &=& A ; \textbf{fail}          & \comment[concatfail2]{/Concat\_fail2/}\\
	 A            &=& A ; \textbf{skip} & \comment[concatskip1]{/Concat\_skip1/}\\
	              &=& \textbf{skip} ; A & \comment[concatskip2]{/Concat\_skip2/}\\
\end{mytable}

The “${\leq}$” and “{$<$}” operators between traces similarly extend to trace sets: A ${\leq}$ B is defined as 

\begin{definitionn}[0.5]{}{And similarly for ``{\textless}''}
$\forall x: \text{A \textbar\ }\exists y:\text{ B \textbar\ }x \leq y$
\end{definitionn}

Unlike the operator \ on traces, “${\leq}$” on trace set is not an order relation since it is not antisymmetric.

We can “slice” trace sets by conditions, pre- and post-. If c is a Boolean expression and A is a set of traces: 

Restriction: only retain traces whose initial state satisfies c

\begin{mytable}
	c / A &${\triangleq}$ \{$x$: A {\textbar} $x_1$ [c]\} && \comment[restrictdef]{[RESTRICT\_DEF]}\\
\end{mytable}

Corestriction: only retain traces whose last state satisfies c

\begin{mytable}
	A {\textbackslash} c &${\triangleq}$ \{$x$: A {\textbar} $x_L$ [c] \} && \comment[corestrictdef]{[CORESTRICT\_DEF]}\\
\end{mytable}

Since we define program semantics by traces, we may use the following notations for programs:

\begin{mydefinition}[traceseteq]{[TRACESET\_EQ]}
	A $\equiv$ B $\coloneqq$ Traces (A) $=$ Traces (B)
\end{mydefinition}
\begin{mydefinition}[tracesetun]{[TRACESET\_UN]}
	A ${\cup}$ B $\coloneqq$ P such that Traces (A) $\cup$ Traces (B) = Traces (P)
\end{mydefinition}
\begin{mydefinition}[tracesetun]{[TRACESET\_SUB]}
A ${\subseteq}$ B $\coloneqq$ Traces (A) $\subseteq$ Traces (B) 
\end{mydefinition}
\begin{definitionn}[0.6]{tracesetmemb}{[TRACESET\_MEMB]}
	$x$ ${\in}$ A $\coloneqq$ $x$ $\in$ Traces (A)
\end{definitionn}


\subsection{Properties of trace sets} \label{properties}

The following theorems (all checked mechanically) express formal properties of trace sets and other basic mechanisms introduced above.

\begin{mytable}
	False / A &$=$ \textbf{fail} && \comment[falserestrict]{/False\_restrict/}\\
	True / A &$=$ A && \comment[truerestrict]{/True\_restrict/}\\
	A {\textbackslash} False &$=$ \textbf{fail} && \comment[falsecorestrict]{/False\_corestrict/}\\
	A {\textbackslash} True &$=$ A && \comment[truecorestrict]{/True\_corestrict/}\\
	c / (d / A) &$=$ (c ${\wedge}$ d) / A && \comment[tworestrict]{/Two\_restrict/}\\
	(A {\textbackslash} c) {\textbackslash} d &$=$ A {\textbackslash} (c ${\wedge}$ d) && \comment[twocorestrict]{/Two\_corestrict/}\\
	(A {\textbackslash} c) ; (d / B) &$=$ (A \textbackslash ~ (c $\land$ d)) ; B) && \comment[corestrictrestrict1]{/Corestrict\_restrict1/}\\
        &$=$ A ; ((c $\land$ d) / B) && \comment[corestrictrestrict2]{/Corestrict\_restrict2/}\\
        &${\subseteq}$ A ; B && \comment[corestrictrestrict3]{/Corestrict\_restrict3/}\\
    (A {\textbackslash} c) ; (${\lnot}$ c / B) &$=$ \textbf{fail} && \comment[corestrictrestrict4]{/Corestrict\_restrict4/}\\
	(A {\textbackslash} c) ; B &$=$ A ; (c / B) && \comment[corestrictrestrict5]{/Corestrict\_restrict5/}\\
%
%
	(v / A) ; B &$=$ v / (A ; B) && \comment[restrictcompose]{/Restrict\_compose/}\\
	A ; (B {\textbackslash} v) &$=$ (A ; B) {\textbackslash} v && \comment[composecorestrict]{/Compose\_corestrict/}\\
	v / (A ${\cup}$ B) &$=$ (v / A) ${\cup}$ (v / B) && \comment[restrictunion]{/Restrict\_union/}\\
	(A ${\cup}$ B) {\textbackslash} v &$=$ (A {\textbackslash} v) ${\cup}$ (B {\textbackslash} v) && \comment[corestrictunion]{/Corestrict\_union/}\\
	A ; (B ${\cup}$ C) &$=$ (A ; B) ${\cup}$ (A ; C) && \comment[composeunion1]{/Compose\_union1/}\\
	(A ${\cup}$ B) ; C &$=$ (A ; C) ${\cup}$ (B ; C) && \comment[composeunion2]{/Compose\_union2/}\\
\end{mytable}

\begin{definitionn}[0.7]{testleq}{/Test\_leq/}
	If t tests A and A ${\leq}$ B, then t tests B.
\end{definitionn}

We will now use these concepts to define programming constructs and what they test.

\subsection{Defining control structures} \label{control}
The standard program instructions and control structures are easy to express as trace sets.

We have already seen \textbf{skip} (defined as the set of one-element traces) and \textbf{fail} (defined as the empty trace set).

Sequencing (block structure) has also been defined already through the operator “+” or its equivalent “;”, which corresponds to the use of this symbol of programming languages. Here it correspondingly concatenates traces.

\subsection{Conditional instructions} \label{conditional}
We define the conditional instruction as 

\begin{definitionn}[0.7]{conddef}{[COND\_DEF]}
	\textbf{if} v \textbf{then} A \textbf{end} ${\triangleq}$ (${\lnot}$ v / \textbf{skip}) ${\cup}$ (v / A)
\end{definitionn}

For the present discussion we will not need the commonly used version of the conditional instruction including an \textbf{else} part, but adding it is trivial.

The definition corresponds to the intuitive semantics of conditionals: an execution of C does nothing if v has value False, and otherwise is an execution of A.

\subsection{The power operator} \label{power}

To define loops in the present formalism, it is useful first to introduce an intermediate mechanism, the repetition, or ``power operator'', applicable to instructions. If \e{A} is an instruction, A$^i$ denotes A iterated $i$ times (\textbf{skip} for \e{$i$ = 0}). The precise definition is by induction:

\begin{mytable}
	A$^0$ &${\triangleq}$ \ \textbf{skip} && \comment[powerbase]{[POWER\_BASE]}\\
	A$^{i+1}$ &${\triangleq}$ \ (A ; A$^i$) && \comment[powerstep]{[POWER\_STEP]}\\
\end{mytable}

\subsection{Loops} \label{loop}
We define the loop instruction $L$, written in programming language notation in the form
\textbf{until} e \textbf{loop} B \textbf{end}, as

\begin{mydefinition}[loopdef1]{[LOOP\_DEF1]}

	$$ L \triangleq \bigcup_{i: ~ \mathbb{N}} (\lnot e ~ / ~ B)^i ~~ \backslash ~~ e$$
\end{mydefinition}

This definition corresponds to the intuitive semantics of loops: an execution of $L$ consists of 0 or more executions of \e{B}, from states in which \e{e} does not hold, such that the last of them produces a state where \e{e} holds. 

Since the definition is a union, we can equivalently replace each element by the union of the preceding ones:

\begin{mydefinition}[loopdef2]{[LOOP\_DEF2]}
	$$L = \bigcup L_i$$
\end{mydefinition}

\noindent
where \ 

\begin{mydefinition}[def2li]{[DEF2\_Li]}
	$$L_i \triangleq \bigcup_{j<i} (\lnot e \ / \ B)^j \ \backslash \ e$$
\end{mydefinition}

L$_i$ describes executions that achieve \e{e} by executing \e{B} repeatedly, but (strictly) less than $i$ times. In particular, L$_0$ is an empty set, meaning \textbf{fail}, and

\begin{mytable}
	L$_1$ &$=$ \textbf{skip} {\textbackslash} e && \comment[loopskip1]{/Loop\_Skip1/}\\
	&$=$ e / \textbf{skip} && \comment[loopskip2]{/Loop\_Skip2/}\\
\end{mytable}

We can also express the L$_i$ sequence (the sequence whose union of all terms defines L) inductively as

\begin{mytable}
	L$_0$ &${\triangleq}$ \ \textbf{fail} && \comment[loop3l0]{/Loop3\_L$_0$/}\\
	L$_{i+1}$ &${\triangleq}$ \ L$_i$ ${\cup}$ ((${\lnot}$ e / B)$^i$ {\textbackslash} e) && \comment[loop3li]{/Loop3\_Li/}\\
\end{mytable}

\subsection{A loop as a recursive conditional}
One way to look at the loop L = “\textbf{until} \e{e} \textbf{loop} \e{B} \textbf{end}” is as a solution to the fixpoint equation

\begin{mytable}
	L &$=$ \textbf{if } ${\lnot}$ \e{e} \textbf{ then } \e{B}; L \textbf{ end} && \comment[fixequa1]{[FIXEQUA\_1]}\\
\end{mytable}

Rather than proving directly that loops as defined above (through the sequence L$_i$) satisfy \referencename{fixequa1}, we consider the following sequence of programs inspired by this equation:

\begin{mytable}
	\underline{L}$_0$ &${\triangleq}$ \ \textbf{fail} && \comment[fixdefbase]{[FIXDEF\_BASE]}\\
	\underline{L}$_{i+1}$ &${\triangleq}$ \ \textbf{if } ${\lnot}$ e \textbf{ then } B; \underline{L}$_i$ \textbf{ end} && \comment[fixdefstep]{[FIXDEF\_STEP]}\\
	&= (e / \textbf{skip}) ${\cup}$ (${\lnot}$ e / (B; \underline{L}$_i$)) && \comment[]{by \referencename{conddef}}\\
	&&&\comment[fixdef2step]{/Fixdef2\_step/}\\
\end{mytable}

It yields a proposed alternative definition L for loops:

\begin{mydefinition}[loopdef3]{[LOOP\_DEF3]}
	$$\underline{L} \triangleq \bigcup_{i:\mathbb{N}} \underline{L_i}$$
\end{mydefinition}

As a reminder, the original definition was (after adaptation) 

\begin{mydefinition}[loopdef2rem]{[LOOP\_DEF2]}
	$$L = \bigcup L_i$$
\end{mydefinition}

with L$_i$ defined by \referencename{def2li} above.

\subsection{The two views are equivalent}
We will now prove that the definitions are equivalent, by showing by induction that L$_i$ = \underline{L}$_i$ for all $i$.

Both L$_0$ and \underline{L}$_0$ are \textbf{fail}. Then for $i \geq 0$: 
\vspace{-0.3cm}
{\rightcomment{/Def\_equiv/}}
\vspace{-0.5cm}
\begin{equation*}
	\setlength{\arraycolsep}{0em}
	\renewcommand{\arraystretch}{1.5}
	\begin{array}{llr}
		L&_{i+1}= \bigcup_{j \leq i} (\lnot e / B)^j \backslash e & \text{\comment{by \referencename{def2li}}}\\
		\underline{L}&_{i+1}= (e / \textbf{skip}) \cup (\lnot e / (B; \underline{L}_i)) & \text{\comment{by \referencename{fixdef2step}}}\\
		&= (e / \textbf{skip}) \cup (\lnot e / (B; L_i)) & \text{\comment{by induction hypothesis}}\\
		&= L_1 \cup (\lnot e / (B; L_i)) & \text{\comment{by \referencename{loopskip2}}}\\
		&= L_1 \cup (\lnot e / (B; \bigcup_{j<i} ((\lnot e / B)^j \backslash e))) & \text{\comment{by \referencename{def2li}}}\\
		&= L_1 \cup ((\lnot e / B) ; \bigcup_{j<i} ((\lnot e / B)^j \backslash e)) & \text{\comment{by \referencename{restrictcompose}}}\\
		&= L_1 \cup \bigcup_{j<i} (\lnot e / B) ; ((\lnot e / B)^j \backslash e) & \text{\comment{by \referencename{composeunion1}}}\\
		&= L_1 \cup \bigcup_{j<i} ((\lnot e / B) ; (\lnot e / B)^j) \backslash e & \text{\comment{by \referencename{composecorestrict}}}\\
		&= L_1 \cup \bigcup_{j<i} ((\lnot e / B)^{j+1}) \backslash e & \text{\comment{by \referencename{powerstep}}}\\
		&= L_1 \cup \bigcup_{1 \leq j \leq i} ((\lnot e / B)^j) \backslash e & \text{\comment{Change of index}}\\
		&= L_1 \cup (L_i - L_1) & \text{\comment{by \referencename{def2li}}}\\
		&= L_i & \text{\comment{("―" is set difference)}} \\
		&& \text{\comment{QED}}
	\end{array}
\end{equation*}


\begin{mydefinition}[underapprox]{/Under\_approx/}
Theorem: L$_i$ \ ${\subseteq}$ \ L for every $i$. (This is also true of \underline{L}$_i$ since it is the same as L$_i$.)
\end{mydefinition}

\subsection{Some consequences}
We call \underline{L}$_i$ the \textbf{i-unrolling} of the loop L. It is of the form

\setlength{\tabcolsep}{0em}
\begin{tabular*}{\columnwidth}{@{}l@{\hspace{0.5em}}l@{\hspace{0.5em}}l@{\hspace{0.5em}}l@{\hspace{0.5em}}l@{\hspace{0.5em}}l@{\hspace{0.5em}}l@{\extracolsep{\fill}}r@{}}
\underline{L}$_i \triangleq$ & \textbf{if} & \multicolumn{6}{l}{\textbf{not} e \textbf{then}} \\
				 &	           & \textbf{if} & \multicolumn{5}{l}{\textbf{not} e \textbf{then}} \\
				 &	           &	         & \multicolumn{5}{l}{B} \\
				 &	           &	         & \textbf{if} & \multicolumn{4}{l}{\textbf{not} e \textbf{then}} \\
				 &             &             &             & \ldots &             &                                  & \\
			     &             &             &             &        & \textbf{if} & \textbf{not} e \textbf{then}     & \\
			     &             &             &             &        &             & B                                & \\
				 &             &             &             &        &             & \textbf{check} False \textbf{end}& \comment{Corresponds to \textbf{fail}} \\
				 &             &             &             &        & \multicolumn{3}{l}{\textbf{end}} \\
				 &             &             &             & \multicolumn{4}{l}{\ldots} \\
				 &             &             & \multicolumn{5}{l}{\textbf{end}} \\
				 &             & \multicolumn{6}{l}{\textbf{end}} \\
				 & \multicolumn{7}{l}{\textbf{end}}
\end{tabular*}

with exactly $i$ occurrences of \e{B} (i.e. if $i$ = 0 the instruction fails, if $i$ = 1 it executes B once or fails, if $i$ = 2 it executes B once or twice or fails etc.).
In the general case, an $i$-unrolling executes B at most $i$ times if it can do so with `e' each time not satisfied, and
otherwise fails.

For every trace x of L, there is a smallest $i$ such that x is a trace of \underline{L}$_i$. By the definitions, x is also a trace of \underline{L}$_j$
for all $j > i$. (Recall that \underline{L}$_i$ ${\subseteq}$ \underline{L}$_j$.)
As a consequence, for every test t of L, there is a minimum $i$ such that t is a test of the i-th unrolling (and all subsequent ones).
This development gives the theoretical framework that we need to unroll loops in the present work's testing strategy. The default unrolling level is 1 (we treat a loop like an \textbf{if … then … end}). The more we unroll, the more extensive the tests will
be.

A “bug” is a test for a specific condition (an incorrectness condition). Note that since an i+1-unrolling includes all the traces, and hence all the bugs, of an i-unrolling, the number of bugs found by a test can only be an increasing function of the unrolling level. (Otherwise, there is something wrong with the implementation of the strategy.)

\section{Implementation: adding loop unrolling to an automated test generation strategy} \label{implementation}
\label{implementation}
\begin{figure}[htbp]
\vspace{-0.3cm}
\centerline{{\includegraphics[width=2.6in]{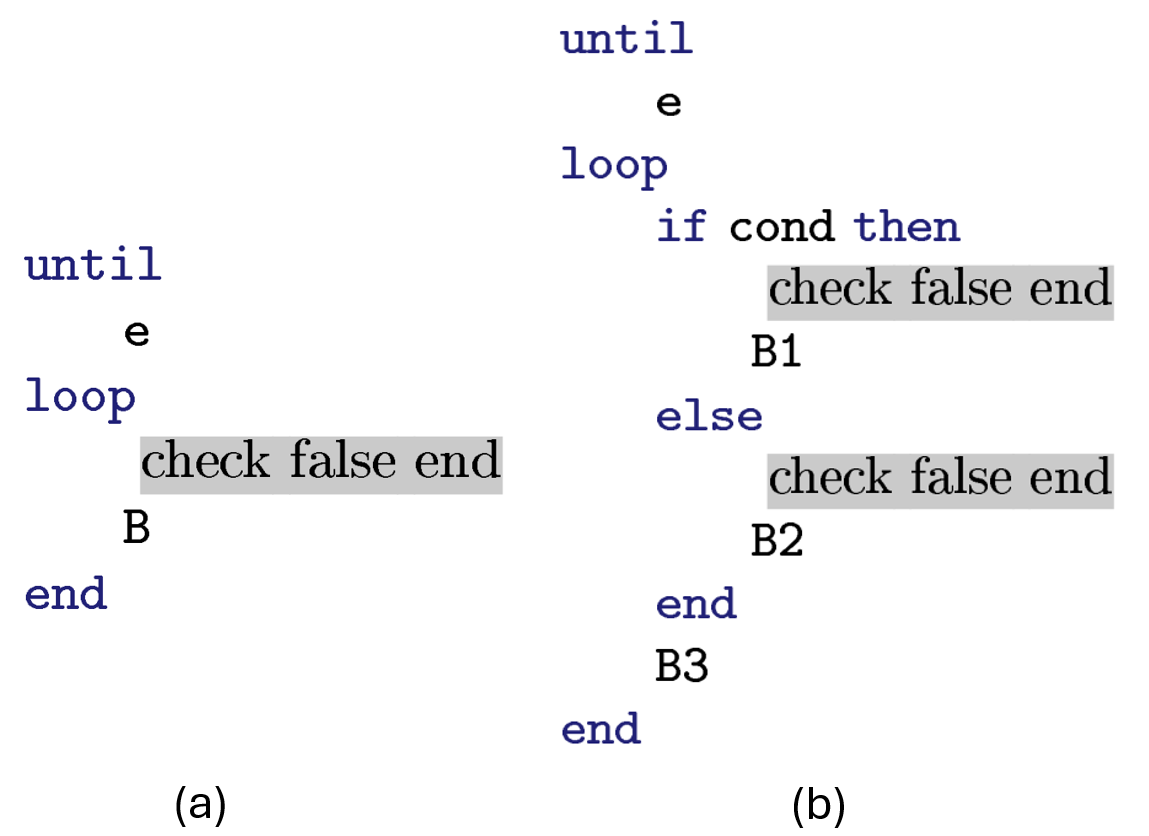}}
}
\caption{Seeding contradictions for a loop}
\label{fig: sc for loop}
\vspace{-0.3cm}
\end{figure}

An automatic test generation strategy called ``seeding contradiction (SC)'', introduced by Huang et al. in \cite{huang2023seeding} (using ideas also applied in other work combining proofs and tests, such as \cite{nilizadeh2022generating}), allows generating test suites that achieve full branch coverage. It relies on the AutoProof tool, a program proving framework internally based on the Boogie prover and an SMT solver such as Z3 \cite{barrett2010smt, de2008z3}.
When generating tests for a loop, the SC inserts a faulty clause in the form of ``\e{check false end}'' (as highlighted in Figure \ref{fig: sc for loop}) inside the loop body. When the loop body contains no conditional statement (a plain block), it injects a contradiction clause at the beginning of the loop body (Figure \ref{fig: sc for loop} (a)); if there are multiple branches inside the loop body, it injects a contradiction clause in each of the branches (Figure \ref{fig: sc for loop} (b)). 

After verifying the seeded version, a program prover will report failures of the seeded clauses (as the assertions will always fail). It obtains a set of counterexample models from the underlying SMT solver, from which it produces executable test cases. Executions of those test cases are guaranteed to go through the locations of the injected ``contradictions'' and thus cover different branches in the loop body.
The SC strategy, however, can only ensure that the generated tests will enter the loop body, without guaranteeing the number of iterations the tests will perform.
To produce tests that will explore the behaviors of the loop to a certain level,
we extend the SC strategy by incorporating loop unrolling. 
We call the extended SC strategy SCU --- seeding contradiction with unrolling.
To allow generating tests that traverse the loop body a specific number of times, SCU performs instrumentation on the code with the loop unrolled to a certain level.

\begin{figure}[htbp]
\vspace{-0.3cm}
\centerline{{\includegraphics[width=2.3in]{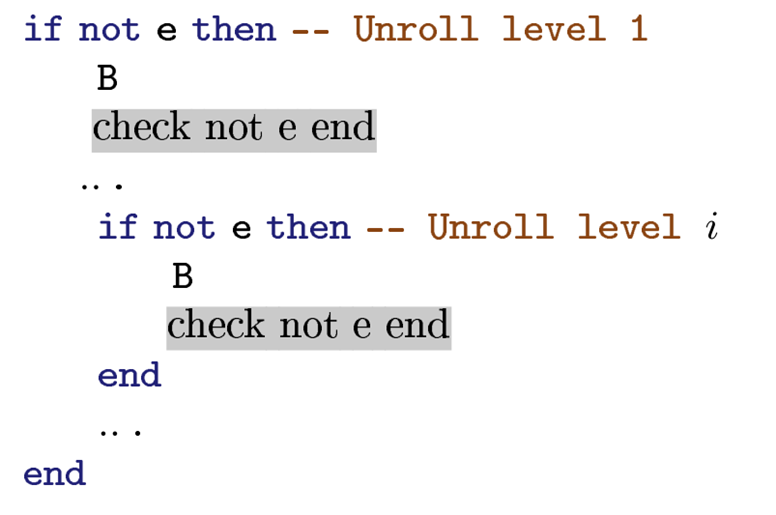}}
}

\caption{SCU for plain loop body}
\label{fig: sc for unrolled plain loop}
\vspace{-0.3cm}
\end{figure}

\begin{figure}[htbp]
\vspace{-0.3cm}
\centerline{{\includegraphics[width=2.3in]{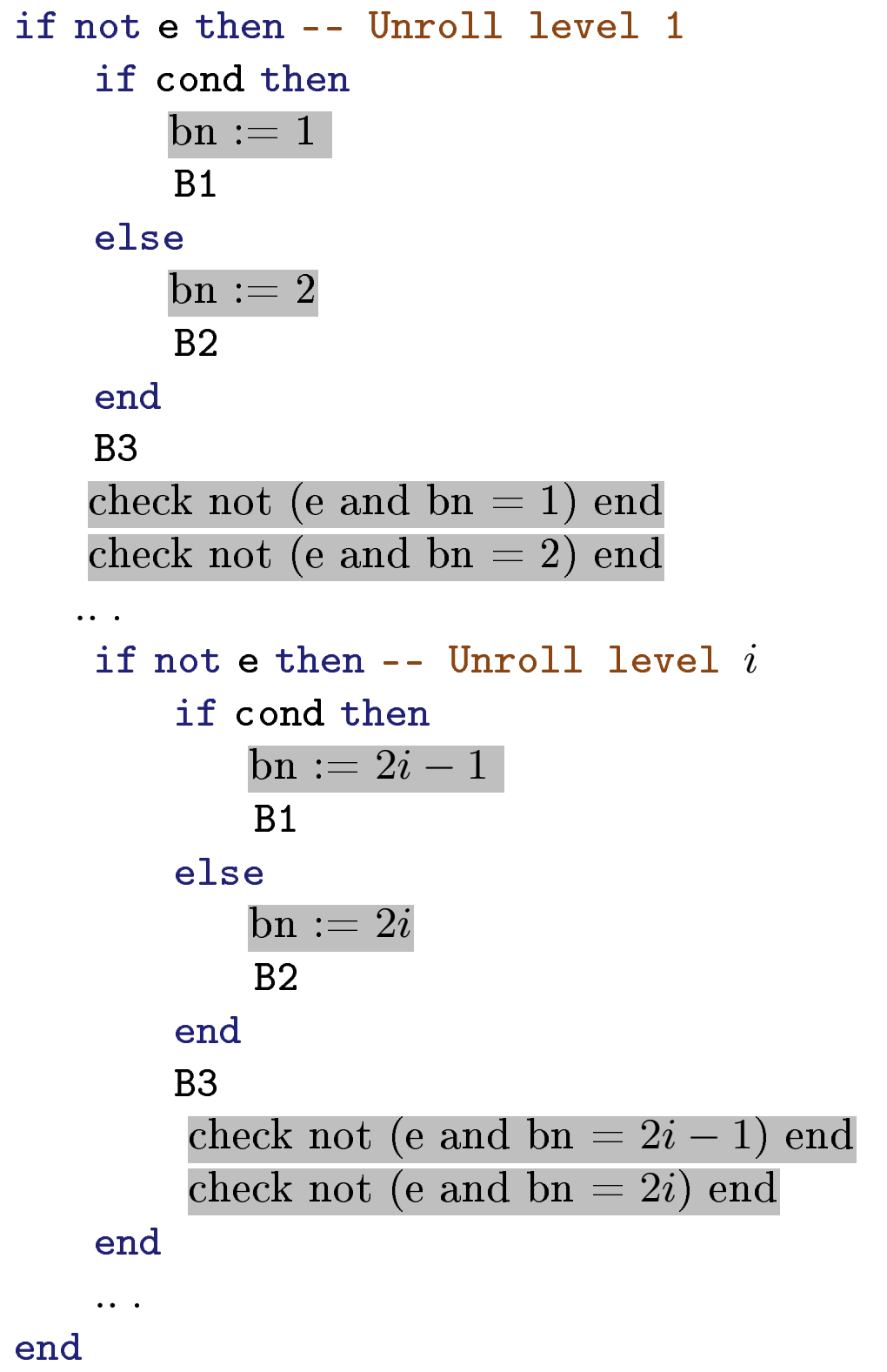}}
}
\caption{SCU for conditionals inside loop body}
\label{fig: sc for unrolled conditional loop}
\vspace{-0.3cm}
\end{figure}

Figure \ref{fig: sc for unrolled plain loop} and \ref{fig: sc for unrolled conditional loop} show the SCU approach. If the loop body is plain, SCU inserts a clause ``\e{check not e end}'' at the end of each unroll level $i\in \{1, ..., n\}$ ($n$ is the loop unrolling factor). 
Adding such an assertion at level $i$ forms a task for the prover --- to find a counterexample for the property ``\e{not e}'' at the end of level $i$. 
If such a counterexample exists, SCU produces a test from the counterexample.
During the execution of the test, the exit condition ``\e{e}'' holds at the end of level $i$, which enforces the loop to exit. In other words, the test is guaranteed to exercise the loop exactly $i$ times.

Verification of the seeded version results in $n$ failures and thus $n$ tests;  one for each unroll level. Note that some of the unrolled levels might not be reachable. For example, when a loop traverses an array whose size should be less than 10, the loop body is unreachable at unrolled level 10 or above. In those cases, the prover will produce no tests for the unreachable levels.

If the loop body contains conditionals, for each unrolling level $i$, SCU produces a test suite consisting of tests that go through every branch. 
Figure \ref{fig: sc for unrolled conditional loop} shows the instrumentation of the unrolled loop, whose body contains two branches.
SCU uses an integer variable \e{bn} to distinguish different branches in the unrolled loop.
Let $m$ be the number of branches in the original loop (here $m = 2$), the value of \e{bn} is in the range $[1, m*n]$.
For each unroll level $i$, it identifies different branches by inserting at each of the branch an assignment ``\e{bn := $j$}'', where $j \in {[m*(i-1) + 1,\ m*i]}$. A unique value $j$ identifies each branch.
At the end of each unroll level, it inserts $m$ assertions in the form of ``\e{check not (e and bn = j)}''. This assertion assigns a task to the prover to find a counterexample that will satisfy the property ``\e{e and bn = j}''.
The test produced from the counterexample is guaranteed to go through the $j^{th}$ branch and establish the exit condition at the end of unroll level $i$.
If all branches are reachable at every unroll level, verification of the instrumented version results in $m*n$ tests.

\section{Evaluation} \label{evaluation}

We evaluate the implementation of SCU and discuss the trade-offs between performance and unrolling depth, with the objective to answer the following research questions:

\begin{itemize}
    \item{\textbf{RQ1}} \emph{What is the precise impact of loop unrolling on test generation time? }
    \item{\textbf{RQ2}} \emph{What is the precise impact of loop unrolling on test execution time?}
    \item{\textbf{RQ3}} \emph{Does loop unrolling actually lead to test suites that find more bugs?}
\end{itemize}

\subsection{Experiment design}
The experiment uses 12 examples that contain loops, adapted from examples in the AutoProof tutorial\footnote{http://autoproof.sit.org/autoproof/tutorial} and benchmarks of previous software verification competitions \cite{weide2008incremental} \cite{bormer2011cost} \cite{klebanov20111st}. Each example contains exactly one loop. 
Table \ref{table: examples} lists their characteristics, including implementation size (number of Lines Of Code), number of branches in the loop body.
\vspace{-0.3cm}
\begin{table}[htbp]
\tiny
    \caption{\textbf{Examples}}
\centering
\scriptsize
 \renewcommand\arraystretch{1.1}
    \begin{tabular}{ p{100pt} p{50pt} p{70pt}}
    \hline
       Example & Lines of Code & \#Branches in the loop
    \\ \hline
    \rowcolor{blue!5} BINARY\_SEARCH & 67 & 3
   \\
   MAX\_IN\_ARRAY & 54 & 2
   \\
  \rowcolor{blue!5} SQUARE\_ROOT & 55 & 3
  \\
  FACTORIAL & 41 & 0
   \\
   \rowcolor{blue!5} GCD (Greatest Common Divisor) & 128 & 2
      \\
    SUM\_AND\_MAX & 57 & 2
      \\
   \rowcolor{blue!5} PRIME\_CHECK & 53 & 2
      \\
    LINEAR\_SEARCH & 45 & 0 \\

   \rowcolor{blue!5} ARITHMETIC\_ADD & 49 & 0
    \\
    ARITHMETIC\_MULTIPLY & 34 & 0
   \\

   \rowcolor{blue!5} ARITHMETIC\_DIVIDE & 32 & 0 \\

    INVERSE & 46 & 2
   \\ \hline
    \end{tabular}%
    \label{table: examples}%
\end{table}

The experiment applies SCU to generate tests for those routines, with loops unrolled to different depths from 1 to 15.
It then compares the fault-identification performance of the resulting test suites.
For each of the examples, the experiment creates different faulty variants by randomly injecting errors into the correctly verified version. To assess the overall performance of each unrolling group, it then performs for each group 20 repetition runs of the following procedures: generate tests; run the tests on the faulty variants; collect the faults found during testing.

All sessions took place on a machine with a 2.1 GHz Intel 12-Core processor and 32 GB of memory, running Windows 11 and Microsoft .NET 7.0.203. Versions used are: EiffelStudio 22.05 (used through AutoProof and AutoTest); Boogie 2.11.10; Z3 solver 4.8.14. All code and results are available at \url{https://github.com/icst-2025-88/loop_unrolling}.

\subsection{Analyses and results}
\noindent \textbf{Impact on Test Generation Time (RQ1)} To answer RQ1: \emph{What is the precise effect of loop unrolling on test generation time?}, we measure the test generation time, including the time for verification and for generating test scripts from counterexamples. Table \ref{table: test generation time} shows the test generation time of SCU when different unrolling depths are applied. 
In some cases, test generation or executing the generated tests would become intractable when unrolling depth exceeds a certain level: the experiment can only handle \e{BINARY\_SEARCH} and \e{SQUARE\_ROOT} up to unroll level 10, and \e{GCD} and \e{PRIME\_CHECK} up to unroll level 8. 


When the depth is below 5, a test generation task costs less than 1 seconds for most of the examples.
For those examples with plain loop body (there is no conditional inside the loop), 
including \e{FACTORIAL}, \e{LINEAR\_SEARCH}, \e{ARITHMETIC\_ADD}, \e{ARITHMETIC\_MULTIPLY}, \e{ARITHMETIC\_DIVIDE}, the overall overhead roughly grows linearly as the unrolling depth increases. The time cost remains at the same scale.
In the other 7 examples with branches inside the loop body, the time cost increases gradually at small depths, but the increment becomes more substantial as the depth becomes larger. 
The difference of the time cost between the initial and the final unrolling depths is significant and occurs on different scales.
Those examples contain at least 2 branches in their loop bodies, resulting in the addition of more contradictory contracts during test generation by SCU.

\begin{table*}[htbp]
    \caption{\textbf{Test generation time}}
\centering
\scriptsize
 \renewcommand\arraystretch{1.1}
    \begin{tabular}{ |p{100 pt} p{20 pt} p{20 pt} p{20 pt} p{20 pt} p{20 pt} p{20 pt} p{20 pt} p{20 pt} p{20 pt} p{20 pt} p{20 pt} p{20 pt} p{20 pt} p{20 pt} p{20 pt} |}
    \hline
       Example & 1 &  2 &  3 &  4 &  5 &  6 &  7 &  8 &  9 &  10 &
        11 &  12 &  13 &  14 &  15
    \\ \hline
    \rowcolor{blue!5} BINARY\_SEARCH & 0.95 & 0.63 & 0.80 & 0.77
& 0.99 & 1.33 & 1.46 & 1.96 & 2.77 & 7.05 & -- & -- & -- & -- & --
   \\
   MAX\_IN\_ARRAY & 0.18 & 0.22 & 0.34 & 0.38 & 0.59& 0.67 &
    0.90& 1.06 & 1.32 & 1.70 & 2.11 & 2.48 &  2.94 & 3.36 & 5.54
   \\
  \rowcolor{blue!5} SQUARE\_ROOT & 0.10 & 0.21 & 0.30 & 0.53 & 0.53 &
    1.16 & 1.67 & 2.97 & 3.85 & 6.41 & -- & -- & -- & -- & --
  \\
  FACTORIAL & 0.38 & 0.10 & 0.11 & 0.13 & 0.13 & 0.15 & 0.16 & 0.19 & 0.21 & 0.24 & 0.27 & 0.30 & 0.33 & 0.35 & 0.40
   \\
   \rowcolor{blue!5} GCD & 0.09 & 0.11 & 0.18 & 0.39 & 0.61 & 0.90 & 1.70 & 4.27 & -- & -- & -- & -- & -- & -- & --
      \\
    SUM\_AND\_MAX & 0.18 &  0.32 & 0.40 & 0.50 & 0.68 & 0.86 &
    1.10 & 1.50 & 1.90 & 2.46 & 2.90 & 3.53 & 3.80 & 4.59 & 6.33
      \\
   \rowcolor{blue!5} PRIME\_CHECK & 0.68 & 0.10 & 0.12 & 0.14 & 0.19 & 0.21 & 0.26 & 0.46 & -- & -- & -- & -- & -- & -- & --
      \\
    LINEAR\_SEARCH & 0.13 & 0.12 & 0.16 & 0.17 & 0.19 & 0.19 & 0.22 & 0.25 &
    0.26 & 0.30 & 0.28 & 0.38 & 0.31 & 0.38 & 0.35 \\

   \rowcolor{blue!5} ARITHMETIC\_ADD & 0.08 & 0.11 & 0.13 & 0.16 &
    0.17 & 0.19 & 0.21 & 0.25 & 0.27 & 0.29 & 0.34 & 0.36 & 0.40 &
    0.42 & 0.46
    \\

    ARITHMETIC\_MULTIPLY & 0.09 &  0.10 & 0.12 & 0.14 & 0.16 & 0.19 &
    0.21  &0.24 & 0.26 & 0.29 & 0.33 & 0.37 & 0.37 & 0.45 & 0.45
   \\

   \rowcolor{blue!5} ARITHMETIC\_DIVIDE & 0.07 & 0.09 & 0.11 &
    0.13 & 0.18 & 0.19 & 0.22 & 0.25 & 0.27 & 0.30 & 0.40 & 0.38 & 0.44 & 0.67 &  0.54 \\

    INVERSE & 0.14 & 0.20 & 0.29 & 0.40 & 0.51 & 0.62 &
    0.79 & 0.97 & 1.12 & 1.27 & 1.48 & 1.68 & 2.0 & 2.32 & 2.57
   \\ \hline
    \end{tabular}%
    \label{table: test generation time}%
\end{table*}

\vspace{0.1cm}
 \noindent \textbf{Impact on Test Execution Time (RQ2)} To answer RQ2: \emph{What is the precise effect of loop unrolling on test execution time?}, we measure the execution time of tests generated with different unrolling depths. Table \ref{table: execution time}
 displays the average execution time for the tests of each unrolling group across the 20 runs.
Overall, the test execution time increases linearly with the unrolling depth, remaining relatively small  --- under 0.5 seconds in most cases. The test execution time for \e{FACTORIAL} is notably high, as its contracts rely on a recursive function, the evaluating the correctness of the contracts incurs additional time cost. 
The increment in test execution time for \e{BINARY\_SEARCH} is more substantial than the others, as it involves an array whose size grows exponentially with the unrolling depth. When the unrolling depth reaches 8, the size of the input array becomes considerably large, requiring more computational resources and time. This results in a significant increase in the execution time when depth rises from depth 8 to 10.
\begin{table*}[htbp]
\vspace{-0.3cm}
    \caption{\textbf{Test execution time}}
\centering
\scriptsize
 \renewcommand\arraystretch{1.1}
    \begin{tabular}{ |p{100 pt} p{20 pt} p{20 pt} p{20 pt} p{20 pt} p{20 pt} p{20 pt} p{20 pt} p{20 pt} p{20 pt} p{20 pt} p{20 pt} p{20 pt} p{20 pt} p{20 pt} p{20 pt} |}
    \hline
       Example & 1 &  2 &  3 &  4 &  5 &  6 &  7 &  8 &  9 &  10 &
        11 &  12 &  13 &  14 &  15
    \\ \hline
    \rowcolor{blue!5} BINARY\_SEARCH & 0.09 & 0.13 & 0.13 & 0.14 &
    0.15 & 0.19 & 0.24 & 0.44 & 1.04 & 3.04 & -- & -- & -- & -- & --
   \\
   MAX\_IN\_ARRAY & 0.01 & 0.02 & 0.02 & 0.03 & 0.04 & 0.05 & 0.05
    & 0.06 & 0.07 & 0.08 & 0.09 & 0.09 & 0.10 & 0.11 & 0.12
   \\
  \rowcolor{blue!5} SQUARE\_ROOT & 0.01 & 0.01 & 0.02 & 0.03 & 0.04 &
    0.04 & 0.05 & 0.06 & 0.06 & 0.07 & -- & -- & -- & -- & --
  \\
  FACTORIAL & 5.99 & 9.55 & 13.06 & 16.65 & 20.16 & 23.72 &
   27.26 & 30.87 & 34.46 & 38.14 & 41.78 & 45.32 & 48.94 & 52.58 & 56.22
   \\
   \rowcolor{blue!5} GCD & 0.01 & 0.01 & 0.02 & 0.03 & 0.04 & 0.04 & 0.05
    & 0.06 & -- & -- & -- & -- & -- & -- & --
      \\
    SUM\_AND\_MAX & 0.01 & 0.01 & 0.02 & 0.03 & 0.04 & 0.04 &
    0.05 & 0.06 & 0.07 & 0.07 & 0.08 & 0.09 & 0.10 & 0.11 & 0.11
      \\
   \rowcolor{blue!5} PRIME\_CHECK & 0.01 & 0.02 & -- & 0.02 & 0.03 &
    0.04 & 0.05 & 0.06 & -- & -- & -- & -- & -- & -- & --
      \\
    LINEAR\_SEARCH & 0.01 & 0.02 & 0.02 & 0.03 & 0.04 & 0.05 & 0.06 &
    0.07 & 0.07 &  0.08 & 0.09 & 0.10 & 0.11 & 0.12 & 0.13
         \\
   \rowcolor{blue!5} ARITHMETIC\_ADD & 0.01 & 0.01 & 0.02 & 0.03 &
    0.04 & 0.05 & 0.06 & 0.06 & 0.07 & 0.08 & 0.08 & 0.09 & 0.10 &
    0.11 &  0.12
    \\
    ARITHMETIC\_MULTIPLY & 0.01 & 0.02 & 0.03 & 0.04 & 0.05 & 0.06 &
    0.06 & 0.07 & 0.08 & 0.09 & 0.10 & 0.11 & 0.12 & 0.13 & 0.14
    \\
   \rowcolor{blue!5} ARITHMETIC\_DIVIDE & 0.01 & 0.02 & 0.02 &
    0.03 & 0.04 & 0.05 & 0.06 & 0.06 & 0.07 & 0.08 & 0.09 & 0.10
    & 0.11 & 0.11 &  0.12
    \\
    INVERSE & 0.01 & 0.02 & 0.02 & 0.03 & 0.04 & 0.05 & 0.05 &
    0.06 & 0.07 & 0.08 & 0.09 & 0.09 & 0.10 & 0.11 & 0.12
   \\ \hline
    \end{tabular}%
    \label{table: execution time}%
\vspace{-0.3cm}
\end{table*}

\vspace{0.1cm}
\noindent \textbf{Effectiveness of SCU (RQ3)} To answer RQ3: \emph{Does loop unrolling actually lead to test suites that find more bugs?}, we run test suites generated with different loop unrolling depth and collect number of detected faults.
A test suite generated in a run may contain several test cases that uncover the same failure (violation of the same contract) multiple times. To avoid the resulting redundancy, the experiment only collects \textit{distinct} faults. A distinct fault is identified by a unique tuple:
\[ <program\ variant,\ tag\ of\ failed\ contract,\ line\ number>\]

The evaluation of the performance of fault detection in each unrolling group of a class $c$ uses the following two criteria:
\begin{itemize}
\item $N_p$: the number of distinct faults detected per run, which can be defined as a function:
\[N_p(i) = \frac{\sum\limits_{j=1}^{20}  |F_c(i, j)|}{20}\]
 where $F (i,\ j)$ is the set of distinct faults found at the $j^{th}$ run ($1 \leq j \leq 20$) with unrolling depth $i$. $N_p (i)$ represents the average performance of fault detection of unrolling group $i$.
 $N_p (i)$ is necessary for the assessment, as different test generation runs use different random seeds (for SMT solving), resulting in different test suites and hence in different detected faults.
 \item $N_a$: the number of all distinct faults that appear during the 20 repetition runs of that group, which can be described as a function over the unrolling depth $i$:
 \vspace{-0.2cm}
\[N_a(i) = |\bigcup\limits_{j=1}^{20} F_c (i,\ j)|\]
\noindent Informally, $N_a (i)$ represents the number of faults that can be found by unrolling group $i$ when the experiment repeats the test generation a sufficient number of times.
\end{itemize}

\noindent Table \ref{table: faults detected per run} and Table \ref{table: faults detected all runs} display the results of $N_p$ and $N_a$ of the 12 examples.
\e{Total} sums up the number of faults of all the examples. 
Overall, the execution of the generated tests detect 251 distinct faults during the experiment.

The result a significant improvement in fault detection as unroll depth increase: an average test suite produced by SCU is able to find 63.7\% ($N_p (1) = 159.75$) of all distinct faults with unrolling depth of 1, while unrolling the loop 5 times effectively uncovers over 80\% ($N_p (5) = 202.85$) of all faults.
Considering all 20 runs, tests generated at unroll depth 1 is able to find 69.3\% ($N_a (1) = 174 $) of all faults; unrolling the loop 5 times is good enough to detect 96.4\% ($N_a (5) = 242$) of the faults.

For most of the examples (10 out of 12), increasing the unrolling depth indeed helps in finding more faults. The benefit is most significant for \e{BINARY\_SEARCH}, \e{GCD} \e{SUM\_AND\_MAX}, \e{SQUARE\_ROOT}, and \e{FACTORIAL}.
For some examples, the benefits brought by unrolling is minimal, typically resulting in only 1 or 2 additional faults detected; those examples include \e{MAX\_IN\_ARRAY},
 \e{ARITHMETIC\_MULTIPLY}, \e{ARITHMETIC\_DIVIDE}, \e{INVERSE}, \e{PRIME\_CHECK}. 
In the cases of \e{LINEAR\_SEARCH} and \e{ARITHMETIC\_ADD}, however, unrolling appears to have no impact, with no additional faults detected as the unrolling depth increases.
Both examples involve straightforward operations within the loop: \e{LINEAR\_SEARCH} iterates through elements, while \e{ARITHMETIC\_ADD} performs simple addition; they only alter the value of just a single variable.
The result suggests that loops with more complex conditional statements or intricate data dependencies are more likely to benefit from unrolling.
In contrast, loops with simple operations like basic summation or iterating through array elements see little advantage from unrolling; the faults in these cases seem identifiable without requiring deeper unrolling.


\begin{table*}[htbp]
    \caption{\textbf{Performance of bug detected per run ($N_p$)}}
\centering
\scriptsize
 \renewcommand\arraystretch{1.3}
    \begin{tabular}{ |p{100 pt} p{22 pt} p{22 pt} p{22 pt} p{22 pt} p{22 pt} p{22 pt} p{22 pt} p{22 pt} p{20 pt} p{20 pt} p{20 pt} p{20 pt} p{20 pt} p{20 pt} p{20 pt} |}
    \hline
       Example & 1 &  2 &  3 &  4 &  5 &  6 &  7 &  8 &  9 &  10 &
        11 &  12 &  13 &  14 &  15
    \\ \hline
    \rowcolor{blue!5} BINARY\_SEARCH & 25.5 & 31.7 & 35.55 & 38.1 & 39.15
    & 39.95 & 41.05 & 41.8 & 42.25 & 42.8 & -- & -- & -- & -- & --
   \\
   MAX\_IN\_ARRAY & 4.9 & 5.25 & 5.45 & 5.45 & 5.5 & 5.5 & 5.55 & 5.55 & 5.55 & 5.6 & 5.65 & 5.7 & 5.8 & 5.8 & 5.8
   \\
  \rowcolor{blue!5} SQUARE\_ROOT & 16 & 18 & 18.85 & 18.95 & 18.95
    & 18.95 & 18.95 & 18.95 & 19 & 19 & -- & -- & -- & -- & --
  \\
  FACTORIAL & 10 & 17 & 17 & 17 & 17 & 17 & 17 & 17 & 17 & 17 & 18 &
    18 & 18 & 18 & 18
   \\
   \rowcolor{blue!5} GCD & 13 & 19 & 19.8 & 19.9 & 19.95 & 20 & 20 & 20 & -- & -- & -- & -- & -- & -- & --
      \\
    SUM\_AND\_MAX & 8.1 & 13.35 & 13.55 & 14.25 & 15.1 & 16.45 & 16.5 & 16.65 & 16.85 & 17.25 & 17.4 & 17.8 & 17.85 & 18.15 & 18.15
      \\
   \rowcolor{blue!5} PRIME\_CHECK & 22 & 22 & 22 & 22 & 22 & 22 & 22 & 22.55 & -- & -- & -- & -- & -- & -- & --
      \\
    LINEAR\_SEARCH & 13 & 13 & 13 &13 & 13 &13 & 13 & 13 & 13 &13 & 13 &13 & 13 & 13 & 13
         \\
   \rowcolor{blue!5} ARITHMETIC\_ADD & 8 & 8 & 8 &8 & 8 &8 & 8 & 8 & 8 &8 & 8 &8 & 8 & 8 & 8
      \\
    ARITHMETIC\_MULTIPLY & 9 & 9.5 & 10 & 10 & 12 & 12 & 12 & 12 & 12 & 12
    & 12 & 12 & 12 & 12 & 12
      \\
   \rowcolor{blue!5} ARITHMETIC\_DIVIDE & 10.5 & 11.5 & 11.5 & 11.5 & 12 & 12 & 12 & 12 & 12 & 12 & 12 & 12 & 12 & 12 & 12
       \\
    INVERSE & 19.75 & 19.95 & 20 & 20.1 & 20.2 & 20.2 & 20.3 & 20.5 & 20.6
    & 20.6 & 20.6 & 20.6 & 20.8 & 20.8 & 20.9
  \\ \hline
  Total & 159.75 & 188.25 & 194.7 & 198.25 & 202.85 &205.05 & 206.35 & 208 & -- & -- & -- & -- & -- & -- & --
   \\ \hline
    \end{tabular}%
    \label{table: faults detected per run}%
\end{table*}

\begin{table*}[htbp]
    \caption{\textbf{Performance of bug detected over all runs ($N_a$)}}
\centering
\scriptsize
 \renewcommand\arraystretch{1.3}
    \begin{tabular}{ |p{100 pt} p{20 pt} p{20 pt} p{20 pt} p{20 pt} p{20 pt} p{20 pt} p{20 pt} p{20 pt} p{20 pt} p{20 pt} p{20 pt} p{20 pt} p{20 pt} p{20 pt} p{20 pt} |}
    \hline
       Example & 1 &  2 &  3 &  4 &  5 &  6 &  7 &  8 &  9 &  10 &
        11 &  12 &  13 &  14 &  15
    \\ \hline
    \rowcolor{blue!5} BINARY\_SEARCH & 34 & 51 & 59 & 64 & 69 & 71 & 72 & 74 & 74 & 74 & -- & -- & -- & -- & --
   \\
   MAX\_IN\_ARRAY & 5 & 6 & 6 & 6 & 6 & 6 & 6 & 6 & 6 & 6 & 6 & 6  & 6 & 6 & 6
   \\
  \rowcolor{blue!5} SQUARE\_ROOT & 16 & 19 & 19 & 19 & 19 & 19 & 19 & 19 & 19 & 19 & -- & -- & -- & -- & --
  \\
  FACTORIAL & 10 & 17 & 17 & 17 & 17 & 17 & 17 &17 & 17 &17 & 18 & 18 & 18 & 18 & 18
   \\
   \rowcolor{blue!5} GCD & 13 & 19 & 20 & 20 & 20 & 20 & 20 & 20 & -- & -- & -- & -- & -- & -- & --
      \\
    SUM\_AND\_MAX & 9 & 16 & 16 & 20 & 20 & 22 & 22 & 22 & 22 & 22 & 22 & 22 & 22 & 22 & 22
      \\
   \rowcolor{blue!5} PRIME\_CHECK & 22 & 22 & 22 & 22 & 22 & 22 & 22 & 23 & -- & -- & -- & -- & -- & -- & --
      \\
    LINEAR\_SEARCH & 13 & 13 & 13 &13 & 13 &13 & 13 & 13 & 13 &13 & 13 &13 & 13 & 13 & 13
         \\
   \rowcolor{blue!5} ARITHMETIC\_ADD & 8 & 8 & 8 &8 & 8 &8 & 8 & 8 & 8 &8 & 8 &8 & 8 & 8 & 8
            \\
    ARITHMETIC\_MULTIPLY & 13 & 14 & 14 &14 & 14 &14 & 14 & 14 & 14 &14 & 14 &14 & 14 & 14 & 14
               \\
   \rowcolor{blue!5} ARITHMETIC\_DIVIDE & 11 & 12 & 12 &12 & 12 &12 & 12 & 12 & 12 &12 & 12 &12 & 12 & 12 & 12
                  \\
    INVERSE & 20 & 22 & 22 &22 & 22 &22 & 22 & 22 & 22 &22 & 22 &22 & 22 & 22 & 22
  \\ \hline
  Total & 174 & 219 & 228 &237 & 242 & 246 & 247 & 250 & -- & -- & -- & -- & -- & -- & --
   \\ \hline
    \end{tabular}%
    \label{table: faults detected all runs}%
\end{table*}

\begin{figure}
    \centering
    \includegraphics[width=3.4in]{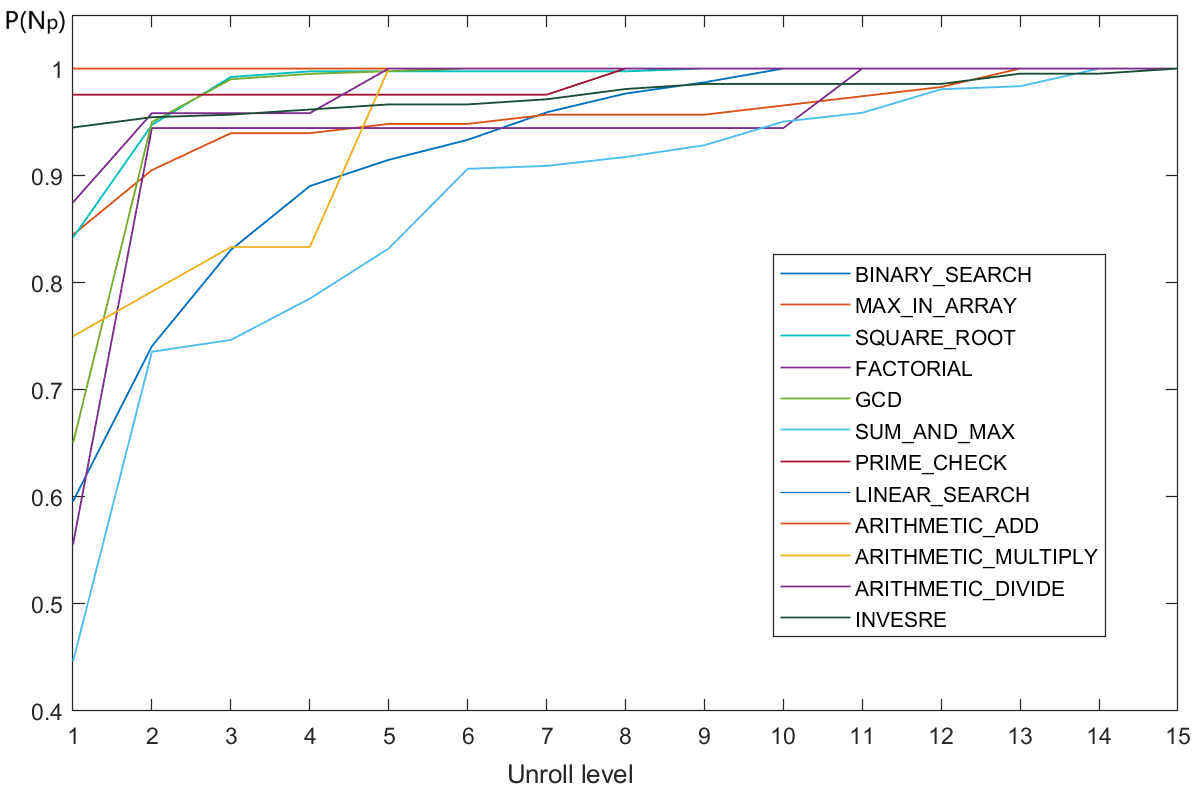}
    \caption{$P (N_p)$: the percentage of faults detected per run at different unrolling levels.}
    \label{figure: percentage of detected faults_all_runs}
\end{figure}
\begin{figure}
    \centering
    \includegraphics[width=3.4in]{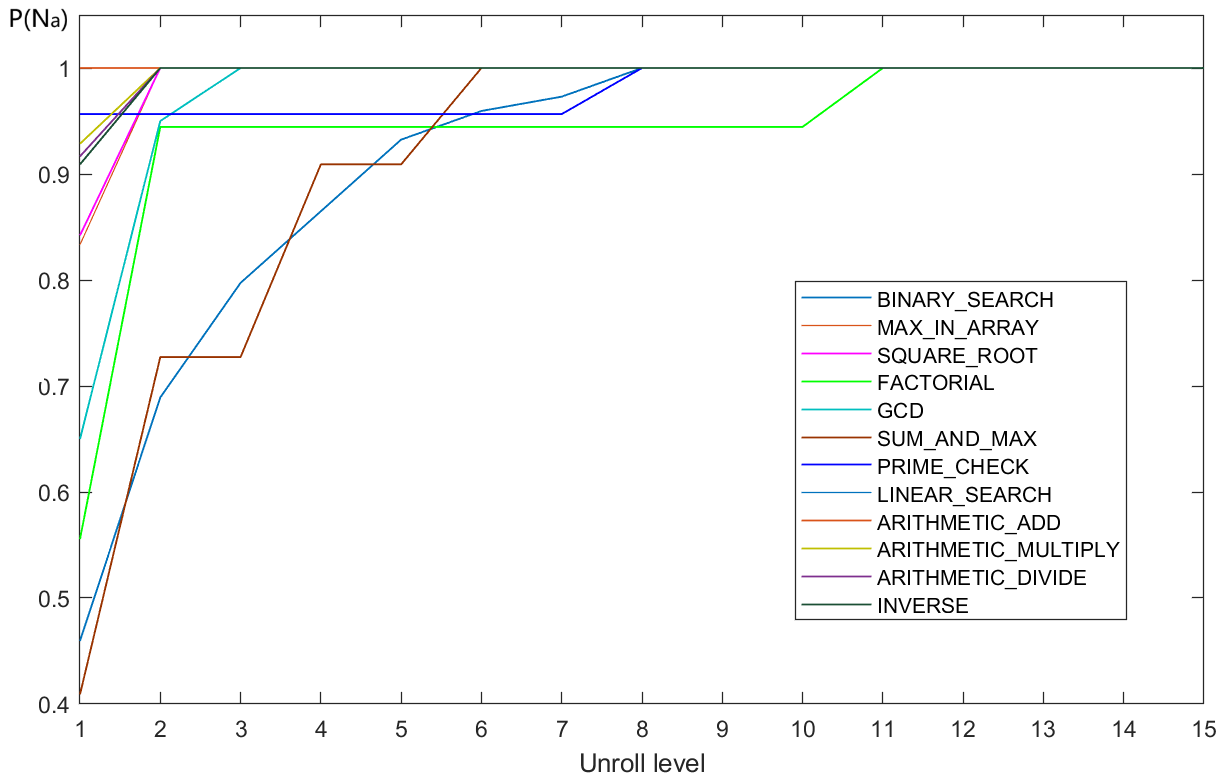}
    \caption{$P (N_a$): the percentage of faults detected over all 20 runs at different unrolling levels.}
    \label{figure: percentage_of detected faults per_run}
\end{figure}

Fig. \ref{figure: percentage_of detected faults per_run} and Fig. \ref{figure: percentage of detected faults_all_runs} depicts the changes of faults detected in percentage when the unroll depth increases. This percentage is computed by dividing $N_p$ and $N_a$ by their maximum values appear during all testing sessions.
The result shows that in most cases, when the depth is small (less than 5), 
the curves rise rapidly, suggesting significant improvements in fault detection.
The most significant improvement occurs when the depth increases from 1 to 2, which results in an improvement of $N_p$ by 11.3\% and $N_a$ by 17.9\%.
As the depth exceeds 5, the effect of unrolling on fault detection diminishes.
Increasing the depth from 5 to 8 yields only a modest improvement of 2.1\% for $N_p$ and 3.6\% for $N_a$, both of which are less substantial than the gains observed at small depths.

The results presented, while still applied to a small set of examples, speak largely for themselves. It is striking to see how much branch coverage --- the gold standard of the testing industry, which in fact often defines a goal of only 80\%, with many teams satisfying themselves with much less --- misses many bugs that unrolling discovers, and continues to discover as the unrolling depth increases.
For automated test generation methods like SC, incorporating unrolling as an essential feature seems crucial. Further research and validation would be beneficial to explore this assertion more thoroughly.

\section{Related work} \label{related}
\label{related_work}
One of the software verification technique that embodies the concept of loop unrolling is
Bounded Model Checking (BMC) \cite{Biere2003BoundedMC},  which reduces model checking of linear temporal logic (LTL) formulas to propositional satisfiability.  
BMC operates by unrolling the transition relation of a finite state machine for a fixed number of steps k, and then checking whether a property violation can occur. If no violation is found, k is increased, and the process is repeated. This approach allows for a systematic exploration of the state space, with the bound k serving as a parameter to control the depth of the search.

The approach of this article, while similar in its unrolling technique, constructs different properties: if a prover finds a counterexample to an assertion, this counterexample serves as a test case, guaranteed to reach both the desired depth in the loop and the specified position within the loop body.
Another distinctive property of the present work lies in its treatment of unwinding correctness. In the BMC literature, the correctness of unwinding is often assumed, either treated as self-evident or accepted axiomatically. The research reported here goes beyond this assumption: we provide a formal proof demonstrating the equivalence between the unrolled loop and the original loop structure.

Two notable tools that implement BMC are CBMC \cite{kroening2014cbmc} and JBMC \cite{brenguier2023jbmc}, which verify C and Java programs against the annotated  assertions, with loops unrolled to a given depth. They can also be used as test generation tool, which automatically generate tests that satisfy a certain code coverage criteria.
This concept of loop unrolling has also been integrated into a test generation tool PathCrawler \cite{williams2005pathcrawler, williams2021towards} which performs unit testing of C programs to obtain path coverage. It reduces the problem of covering all loop paths in a loop is to a \emph{k-path} objective with the aim of covering loop paths within k loop iterations. Compared to the present work, such tools apply the idea of loop unrolling to either improve the efficiency of exploration of systems' behaviors or to systematically cover different loop paths.

Like the present work, Huster et al. \cite{huster2015efficient} went beyond the traditional criterion of branch coverage and proposed an approach to detect more possible failures by explicitly addressing various patterns of loop iteration orders. They group iteration orders that influence one another into equivalence classes based on how the current loop iteration affects the next, thereby reducing the complexity of covering all possible loop path variations.

\section{Threats to validity and open issues} \label{threats}

A limitation of the present work is the small size of the sample set of programs, and the small size of these programs themselves. Although some of them are real software elements (extracted from widely used libraries), they are not representative of large-scale production programs. They do, however, include significant loops, some of them sophisticated, and so provide a credible basis for studying the potential effects of unrolling.

Another issue is that  many of the bugs (although not all) are seeded, rather than being actual bugs found in released code. 

Also limiting the generalization of the present results is the use of an automated verification framework, AutoProof and the associated ``seeding contradiction'' test-generation framework of Huang et al., which at this stage is still a research tool rather than a deployed production environment.

While the results obtained in the  experiments reported above seem strongly to suggest that loop unrolling may be feasible without an undue effect on testing time, they do not clearly uncover a ``magic unrolling number'' --- an absolute constant \e{N} which would enable us to give a general rule-of-thumbs advice to practicing software developers, as in ``unroll 5 times and you will be OK most of the time''. Looking at Fig. \ref{figure: percentage_of detected faults per_run} and Fig. \ref{figure: percentage of detected faults_all_runs} does suggest that something around \e{N = 5} would make sense, but one would need numerous experiments on large and representative code examples before such an initial heuristic would have enough confirmation to warrant inclusion into standard industry guidelines. We hope that the present work provides a solid basis for performing such experiments leading to firm empirical conclusions.

\section{Conclusion} \label{conclusion}

This paper has pursued both a theoretical aim and a practical one. The theoretical contribution is to provide a simple and sound mathematical theory of loop unrolling, avoiding the ambiguities and confusions that may result from a purely informal approach; all the corresponding properties come with a mechanically-checked proof with a leading proof tool, Isabelle. The practical contribution is to apply this theory to generate loop-unrolled test suites for a number of examples, of which some involve sophisticated loops. With the qualifications given in the previous section, the results indicate that: (1)  Standard branch coverage, by ignoring loop body repetitions, \textit{does} miss a significant number of bugs. (2) It \textit{is} practically possible to correct this deficiency by adding automatically unrolled versions of loops to a test suite. (3) For small unrolling levels, the time penalty is reasonable, and justified by the potential for finding extra bugs. (4)  The clear bug-finding outcomes confirm the usefulness of using a formal verification framework and applying it to generate high-coverage test suites, as exemplified in recent research efforts at the frontier of tests and proofs, two complementary techniques of program verification.
    

We hope that these steps provide important information on an essential, if often overlooked, issue of software engineering: is the cavalier attitude that the industry commonly applies to loops, by ignoring the property that actually \textit{defines} the concept of loop (the ability to  repeat instructions!), justified --- and should we not do better?

	
\bibliographystyle{splncs04}
\bibliography{reference}

\end{document}